# Web Based Cross Language Plagiarism Detection

*Chow Kok Kent, Naomie Salim*

*Faculty of Computer Science and Information Systems, University Teknologi Malaysia, 81310 Skudai, Johor, Malaysia*

**Abstract**— As the Internet help us cross language and cultural border by providing different types of translation tools, cross language plagiarism, also known as translation plagiarism are bound to arise. Especially among the academic works, such issue will definitely affect the student's works including the quality of their assignments and paper works. In this paper, we propose a new approach in detecting cross language plagiarism. Our web based cross language plagiarism detection system is specially tuned to detect translation plagiarism by implementing different techniques and tools to assist the detection process. Google Translate API is used as our translation tool and Google Search API, which is used in our information retrieval process. Our system is also integrated with the fingerprint matching technique, which is a widely used plagiarism detection technique. In general, our proposed system is started by translating the input documents from Malay to English, followed by removal of stop words and stemming words, identification of similar documents in corpus, comparison of similar pattern and finally summary of the result. Three least-frequent 4-grams fingerprint mathching is used to implement the core comparison phase during the plagiarism detection process. In K-gram fingerprint matching techinique, although any value of K can be considered, yet K = 4 was stated as an ideal choice. This is because smaller values of K (i.e., K = 1, 2, or 3), do not provide good discrimination between sentences. On the other hand, the larger the values of K (i.e., K = 5, 6, 7...etc), the better discrimination of words in one sentence from words in another. However each K-gram requires K bytes of storage and hence space-consuming becomes too large for larger values of K. Hence, three least-frequent 4-grams are the best option to represent the sentence uniquely.

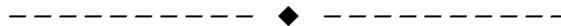

## 1 INTRODUCTION

Academic plagiarism is defined as inappropriate use by student of material originated by others. The legal and ethical frameworks are far from clear, and the Internet has brought new opportunities for plagiarists and new investigative tools for their opponents. Numerous studies show that plagiarism and other types of academic fraud is increasing among students. The most significant phenomenon among the students is during their thesis writing period. In a recent article published by the Center for Academic Integrity (CAI), Professor McCabe claims that "On most campuses, 70% of students admit to some cheating" while "Internet plagiarism is a growing concern" because although only "10% of students admitted to engaging in such behavior in 1999, almost 40%" admitted to it in 2005" [1].

In general, there are various forms of plagiarism such as straight plagiarism, simple plagiarism with footnote, complex plagiarism using footnote, plagiarism using citation but without quotation marks, and paraphrasing as plagiarism. The practice of plagiarism is a form of academic high treason because it undermines the entire scholarly enterprise. Plagiarized news, magazines articles and web resources are the area of concern in this plagiarism issue.

Translation plagiarism is considered as translation of a sentence in the source language in the way that ends up using almost the exact same words as the original source used by the original author [2]. Even though translation plagiarism does not copy the original work directly, it still considered as plagiarism as it try to imitate and use other people ideas or thoughts without any citation or quotation marks. These incidents are much harder to detect, as translation is often a fuzzy process that is hard to search for, and even harder to stop as they usually cross international border [3].

Several arguments including Intellectual Property (IP), ethics, legal constraint, and copyright has been raised up due to the plagiarism issue. Intellectual property (IP) is a legal property right over the creation of the mind, both artistic and commercial, and the corresponding fields of law [4]. Where argument is provided, it is often along the lines that plagiarism is morally wrong because the plagiariser both claims a contribution that they are not justified in claiming, and denies the originator the credit due to them. One further body of law requires more detailed consideration. Copyright law provides the author or commissioner of a work with a small basket of specific rights in relation to the work.

## 2 RELATED WORKS

Here are a few existing plagiarism detection tools or services that assist in plagiarism detection [5].

**Turnitin.com (formerly Plagiarism.org)**
http://www.turnitin.com

Turnitin.com uses digital fingerprinting to match submitted papers against internet resources and an in-house database of previously submitted papers. According to Satterwhite and Gerein, Turnitin.com has the highest rate of detection amongst subscription detection tools. Papers can be submitted individually by either students or lecturers. All papers are archived for future checking – a feature which is particularly useful if copying of previous students' papers is suspected. [6]

**Plagiserve.com http://www.plagiaserve.com**

Plagiserve.com is a free service which searches the internet for duplicates of submitted papers, analyses them, and provides evidence of plagiarism to the lecturer. It has an extensive database of 90,000 papers, essays and CliffNotes study guides, and papers from all known paper mills. Reports are generated in 12 hours. The service is only available through its website, and papers must be submitted in one batch.

**CopyCatch Gold http://www.copycatch.freeserve.co.uk**

CopyCatch Gold is a stand-alone desktop software which can be either installed on a single PC or on a network. It detects collusion between students by checking similarities between words and phrases within work submitted by one group of students. One of the special features of CopyCatch is that it assists students in their writing development, by allowing them to see where they are repeating text from other sources and from their own previous assignments. Its weakness is the inability to detect material downloaded from the web.

**Eve2 – Essay Verification Engine**
http://www.canexus.com/eve/index.shtml

Eve2 is a windows based system, installed on individual workstations. It is not easily installed on servers. Papers are submitted by cutting and pasting plain text, Microsoft Word, or Word Perfect documents into a text box. The program then searches internet resources for matching text. Reports are provided within a few minutes, highlighting suspect text, and indicating the percentage of the paper that is plagiarised. Identified limitations include the inability to trace documents that are not in html format, inability to trace collusion between students, the inability to search subscription websites.

**WordCheck Keyword DP**
http://www.wordchecksystems.com/wordcheck-dp.html

WordCheck Keyword DP is a software program which identifies key word use, and matches documents based on word use and frequency patterns. Reports showing key word profiles and word frequency lists are generated. Limitations of the program include the time-consuming manual checking of each document and the inability of the program to detect plagiarism from internet sources.

Most of the detection softwares mentioned above use the well-known journal or papers database such as ProQuest®, FindArticles®, and LookSmart® as their corpus. Although these tools provide an effective alternative in plagiarism detection, the tools not derived for translated plagiarism. For instance, if someone translates exactly a copy of English paper into Bahasa Melayu, no tool will ever detect the translated version as plagiarized

## 3 OPERATIONAL FRAMEWORK

In this paper, we focus on detecting the Malay-English plagiarism. As a web based plagiarism detection system, our corpus is built up with most of the Internet resources that are detectable by the Google search engine.

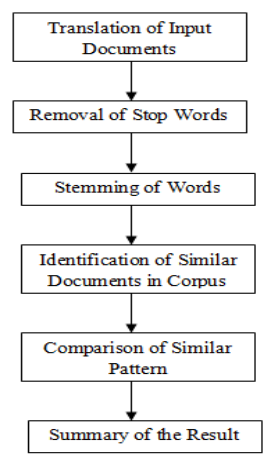

Fig.1 Cross Language Plagiarism Detection Framework

### 3.1 Translation of Input Documents

In order to detect translation plagiarisms, it is essential to translate the plagiarized Malay documents into English before used as the query documents for further detection process. After the plagiarized documents have been translated into English, it will improve the effectiveness of the detection process as the source documents are also in English. We use Google Translate API which is a well-known translation tool developed by the Google and is freely distributed. With this API, the language blocks of text can be easily detected and translated to other preferred languages. The API is designed to be a simple and easy to detect or translate languages when offline translation is not available.

### 3.2 Removing Stop Words

Stop words are the words that frequently occurred in documents. These words do not give any hint values or meanings to the content of their documents, hence they are eliminated from the set of index terms [7]. Salton and McGill (1993) reported that such words comprise around 40 to 50% of a collection of documents text words [8]. Eliminating the stop words in automatic indexing will speed the system processing, saves a huge amount of space in index, and does not damage the retrieval effectiveness [9].

In this paper, before passing the translated documents for comparison through the Internet, it is essential for us to remove the stop words in the translated text. English stop words will be removed in the translated texts. Currently, there are several English stop words list that commonly used in the information retrieval process. Some of the general English stop words are shown as below.

a, an, the, ourselves, been, anywhere, any, by, did, each, ever, even, would, could, few, than, all

Before removing stop words:
*In information retrieval, stop words are the words that frequently occurred in the documents.*
After removing stop words:
*information retrieval, stop words words frequently occurred documents.*

### 3.3 Stemming Words

Stemming is a process to remove the affixes (prefixes and suffixes) in a word in order to generate its root word. Using root word in pattern matching provides a much better effectiveness in information retrieval. There are many stemmers available for the English language such as Nice Stemmer, Text Stemmer and Porter Stemmer are the well-known English stemmer that commonly been used.

Affix stripping is one of the common algorithms used in stemming process. Affix stripping includes the removal of prefixes and suffixes of a word. It does not rely on a lookup table that consists of inflected and roots form. Instead, it is a smaller list of rules that provide the core of the algorithm.

Prefix Stripping:
*If the word begin in "in", remove the 'in'*
*If the word begin in "inter", remove the 'inter'*

Suffix Stripping:
*If the word ends in 'ed', remove the 'ed'*
*If the word ends in 'ing', remove the 'ing'*
*If the word ends in 'ly', remove the 'ly'*

In this paper, we propose the use of Porter Stemming algorithm in our stemming process. The Porter stemming algorithm (or 'Porter stemmer') is a process for removing the commoner morphological and inflexional endings from words in English. Its main use is as part of a term normalisation process that is usually done when setting up information retrieval systems. The original stemming algorithm paper was written in 1979 in the Computer Laboratory, Cambridge (England), as part of a larger IR project, and currently is widely used as a stemming algorithm which is fully tested for its accuracy and effectiveness.

### 3.4 Identifying Similar Documents in Corpus

Corpus (collection of documents) can be either intra-corpus or inter-corpus. Intra-corpus is defined as a collection of documents which are not distributed over the heterogeneous network and can be found in the same storage. Inter-corpus is the collection of documents that located around the World Wide Web.

Instead of using an intra corpus, it is preferable to use a inter corpus which consists of a collection of sources through the Internet. In this case, search engine is an effective alternative. An internet search engine can be used to look for certain keywords and key sentences from a suspected document on the World Wide Web. The detection process becomes more effective as the World Wide Web acts as a large collection of document (corpus) and enables small and characteristic fragments translation. Query documents or texts are inserted into the search engine and a set of results are generated by the search engine. These results show the similar part of the query documents and their corresponding similar sources of documents.

In this paper, we propose the use of Google AJAX Search API (Google search engine) as our corpus. Google search engine is considered as one of the most complete online resources collection. Even papers published in some free server or outside the well-known journal database are also detectable using the Google search engine. The stemmed plagiarized documents are inserted into the search engine and a set of similar source documents can be retrieved.

### 3.5 Comparison of Similar Pattern

Fingerprint matching technique is widely used in the plagiarism detection tools. Fingerprinting divides the document into grams of certain length k. In full fingerprinting, document fingerprint consists of the set of all possible substrings of length K. The fingerprints of two documents can be compared in order to detect plagiarism. In this paper, we represent the fingerprints of each statement in the documents by three least-frequent 4-grams. Although any value of K can be considered, yet K = 4 was stated as an ideal choice by Yerra and Ng (2005) [10]. This is because smaller values of K (i.e., K = 1, 2, or 3), do not provide good discrimination between sentences. On the other hand, the larger the values of K (i.e., K = 5, 6, 7...etc), the better discrimination of words in one sentence from words in another. However each K-gram requires K bytes of storage and hence space-consuming becomes too large for larger values of K.

Three least-frequent 4-grams are the best option to represent the sentence uniquely. To illustrate the three least-frequent 4-gram construction process, consider the following sentence S *"soccer game is fantastic"*. The 4-grams are *socc, occe, ccer, cerg,* etc. In this method, instead of comparing all possible 4-grams, only three 4-grams which have the least frequency over all 4-grams will be chosen.

Let the document contain J distinct n-grams, with $m_i$ occurrences of n-gram number $i$. Then the weight assigned to the $i^{th}$ n-gram will be

$$x_i = \frac{m_i}{\sum_{j=1}^{J}(m_j)}$$

Where

$$\sum_{i=1}^{J}(x_i) = 1$$

The three least-frequent 4-grams are concatenated to represent the fingerprint of a sentence from the query document to be compared with the three least-frequent 4-gram representations of sentences in the source documents. Finally, two sentences are treated the same if their corresponding three least frequent 4-gram representations are the same. A measure of resemblance for each pair of documents is computed as follows:

$$R = \frac{F(A) \cap F(B)}{F(A) \cup F(B)}, 0 \leq R \leq 1$$

where F(A) and F(B) are the common fingerprints in documents A and B, respectively.

### 3.6 Summary of the Result

After gathering the result, a summary of the plagiarism detection is displayed. The plagiarized parts of the documents, source of the plagiarism, and the similarity percentage are all shown in the summary. Further works are needed in order to determine the existence of proper citation or quotations in those plagiarized documents.

## 4 DISCUSSION AND FUTURE WORKS

### Volume of the Corpus

Plagiarism detection tools concentrate on the speed and width of detection, at the cost of quality of detection. The speed is the processing time and the respond time of the system once the data is inserted. The detection width refers to the database available for searching the source document given the suspected plagiarized documents. Hence, the probability in detecting the plagiarism is higher if the volume of the corpus is larger. However, difficulties arise when searching similar document in some small sites where authors who have published their paper in a journal and keep their own copy in their own server for free. Their papers are not published in some well-known journal or papers database such as ProQuest®, FindArticles®, and LookSmart®. In this case, the papers are hardly detectable using own developed search engine.

### Paraphrasing Issue

Paraphrasing is the techniques to modify the structure of an original sentence by changing the sentence's structure or replaces some of the original words with its synonym. Without any proper citation or quotation marks, it also considered as plagiarism. However, it is generally believed nowadays that fingerprint-based approach is quite weak since even slight textual modification can considerably affect the fingerprint of the document. Most of the plagiarism tools are not stable against synonyms. If someone copies and systematically changes words by using synonyms or such, all plagiarism tools we are aware of will fail.

### Authorship Identification and Stylometry Analysis

Authorship identification and stylometry analysis can be used as an alternative in detecting plagiarism indirectly. It can be an indication of plagiarism if there is no essay chunk can be found either on Internet or inside other stu-dent's submission. . If the suspected documents consist of a large amount of text, stylometry analysis will be an effective detection techniques where each parts of the documents can be investigated based on the authorship and style of writing.

## 5  CONCLUSION

In this paper we presented our web based cross language plagiarism detection. Translation plagiarism is becoming a major issue and concern especially in the academic works. With our experiment, we feel that our system can detect the translation plagiarism with high efficiency and effectiveness. Google API can be further utilized in our detection system to improve the detection performance.

## 6  ACKNOWLEDGEMENT

This project is sponsored partly by the Ministry of Science, Technology and Innovation under the E-Science grant 01-01-06-SF-0502.

.

**Mr. Chow Kok Kent** received his B.Sc. degree in Universiti Teknologi Malaysia, Malaysia in 2009. He is currently purchasing Master degree in Faculty of Computer Science and Information System, Universiti Teknologi Malaysia. His current research interest includes information retrieval, Plagiarism Detection and Soft Computting.

**Dr. Naomie Salim** is an Assoc.Prof. presently working as a Deputy Dean of Postgraduate Studies in the Faculty of Computer Science and Information System in Universiti Teknologi Malaysia. She received her degree in Computer Science from Universiti Teknologi Malaysia in 1989. She received her Master degree from University of Illinois and Ph.D Degree from University of Sheffield in 1992 and 2002 respectively. Her current research interest includes Information Retrieval, Distributed Database and Chemoinformatic.

.